\documentclass{PoS}

\usepackage{graphicx}
\usepackage{amsmath}
\usepackage{bm}
\usepackage{amsfonts}

\newcommand{\psib}{{\overline{\psi}}}

\newcommand{\beq}{\begin{equation}}
\newcommand{\eeq}{\end{equation}}
\newcommand{\beqa}{\begin{eqnarray}}
\newcommand{\eeqa}{\end{eqnarray}}

\title{The generalized fermion-bag approach}

\ShortTitle{The generalized fermion-bag approach}

\author{Shailesh Chandrasekharan \textnormal{and} \speaker{Anyi Li} \footnote{Address since September 2011 : Institute for Nuclear Theory, University of Washington, Seattle, 98195, USA}\\
        Department of Physics, Box 90305, Duke University, Durham, North Carolina 27708, USA\\
        E-mail: \email{sch@phy.duke.edu} and \email{anyili@phy.duke.edu}}
        
\abstract{We present a new approach to some four-fermion lattice field theories which we call the generalized fermion bag approach. The basic idea is to identify unpaired fermionic degrees of freedom that cause sign problems and collect them in a bag. Paired fermions usually act like bosons and do not lead to sign problems. A resummation of all unpaired fermion degrees of freedom inside the bag is sufficient to solve the fermion sign problem in a variety of interesting cases. Using a concept of duality we then argue that the size of the fermion bags is small both at strong and weak couplings. This allows us to construct efficient algorithms in both these limits. Using the fermion bag approach, we study the quantum phase transition of the 3D massless lattice Thirrring model which is of interest in the context of Graphene. Using our method we are able to solve the model on lattices as large as $40^3$ with moderate computational resources. We obtain the precise location of the quantum critical point and the values of the critical exponents through this study. }

\FullConference{ The XXIX International Symposium on Lattice Field Theory - Lattice 2011\\
July 10-16, 2011\\
Squaw Valley, Lake Tahoe, California}

\begin{document}

\section{Introduction}

Monte Carlo methods for lattice field theories with massless fermions in three or more dimensions continue to pose a variety of challenges. For example, the popular Hybrid Monte Carlo (HMC) method encounters many difficulties. Sometimes the method encounters sign problems, while in other cases small eigenvalues of the fermion matrix lead to severe singularities. For this reason most practical calculations have always relied on extrapolations to the massless limit. Studies in Quantum Chromodynamics have shown that reliable extrapolations require calculations at many small fermion masses~\cite{Sharpe:2007yd}. As far as we know Monte Carlo calculations on large lattices with exactly massless fermions have never been performed so far with traditional Monte Carlo methods including the HMC. 

Recently, an alternative Monte Carlo method called the fermion bag approach was proposed to solve some four-fermion lattice field theories with exactly massless fermions~\cite{Chandrasekharan:2009wc}. It was shown that the method is extremely efficient at strong couplings where traditional Monte Carlo methods fail. Here we argue that the method is also quite general and can be applied to a variety of problems and in addition, contains an interesting strong-weak coupling duality which makes the method efficient even at weak couplings. The efficiency is due to the fact that the required effort to perform a single local update scales like the square of the number of fermion degrees of freedom inside a bag instead of the space-time volume. Interestingly, the bag size is a small fraction of the thermodynamic volume at both strong and weak couplings. In the massless lattice Thirring model which we study here as a first application, the fermion bags typically contain only an eighth of the total degrees of freedom even at the quantum critical point. Due to this feature, for the first time we are able solve a three dimensional lattice field theory containing exactly massless fermions close to an interesting quantum critical point on lattices as large as $40^3$ with modest computing resources. Through a careful finite size scaling analysis we are able to extract the critical exponents accurately at the quantum critical point in this model. 

\begin{figure*}[t!]
\begin{center}
\hbox{
\includegraphics[width=2.9in]{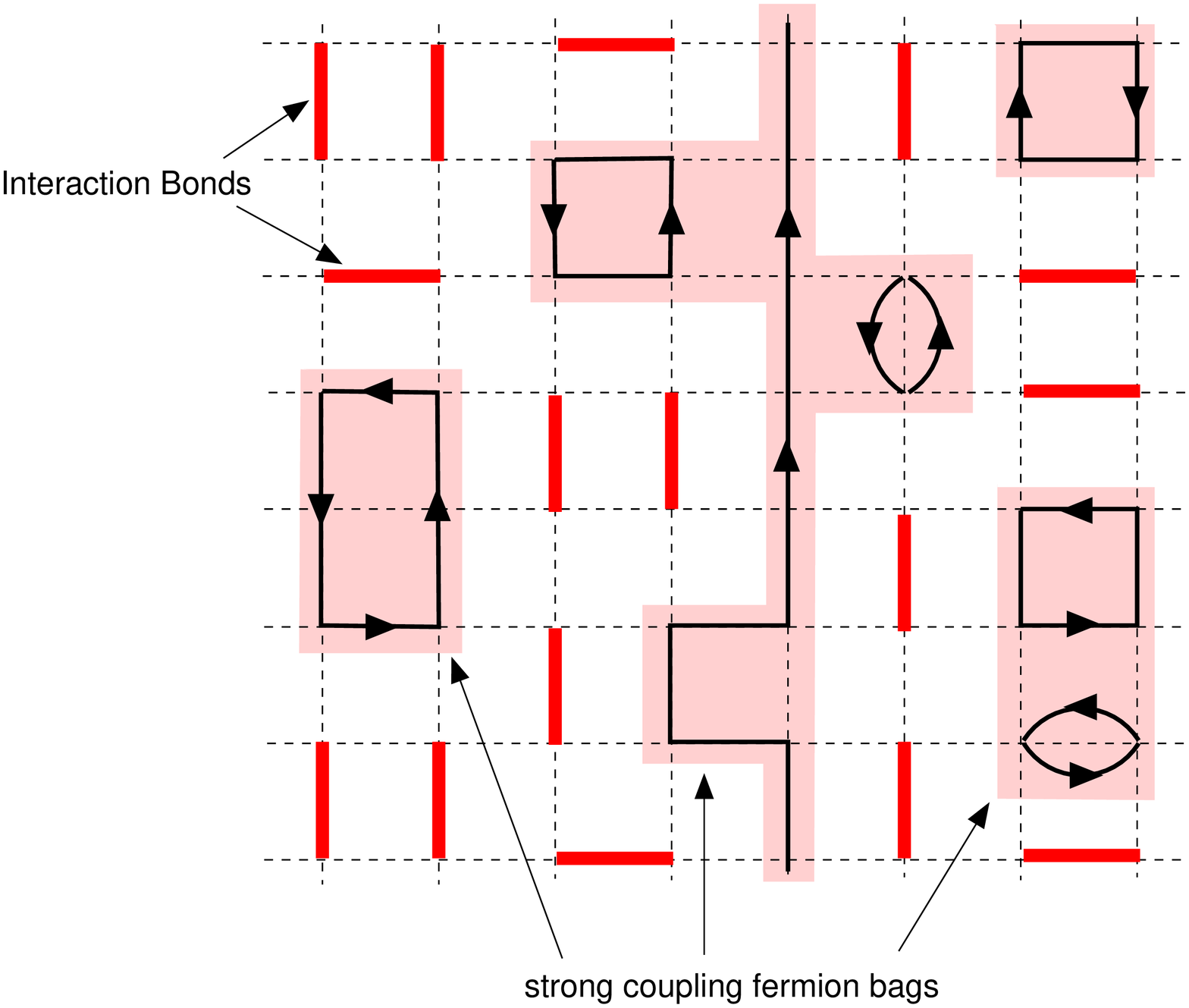}
\hskip0.3in
\includegraphics[width=2.95in]{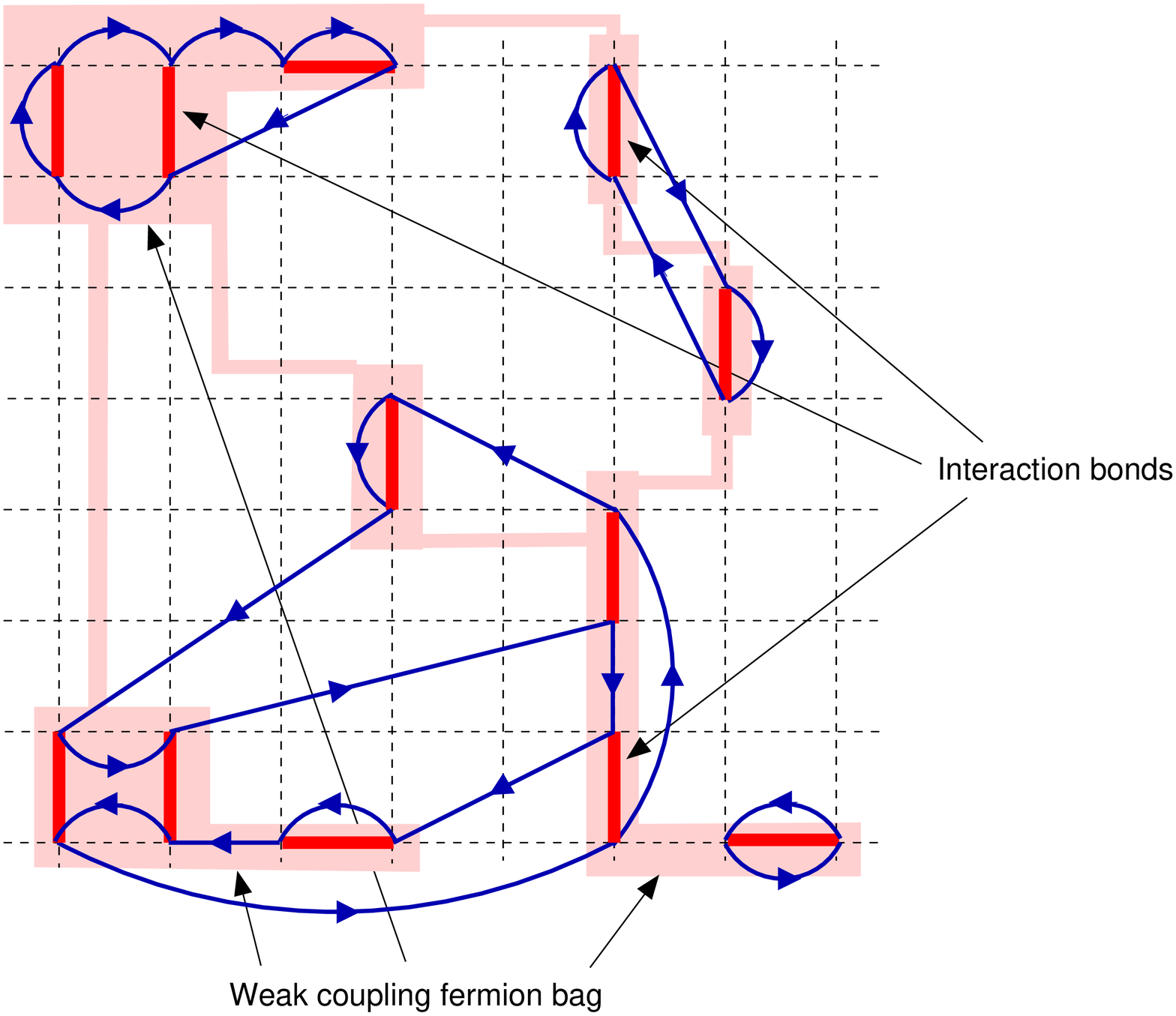}
}
\end{center}
\caption{\label{fig:bag} An illustration of a ``fermion-bag'' configuration at strong couplings (left) and weak couplings (right). The interaction sites are represented through sold bonds and the fermion bags are represented through the shaded region. At strong couplings the fermion bag are made up of free sites and breaks up into many disconnected pieces as seen in the left figure, while at weak couplings the bag contains interaction sites.}
\end{figure*}

\section{The Fermion Bag Idea}

The idea behind the fermion bag is to identify fermion degrees of freedom that cause sign problems and collect them in a bag and sum only over them. This is in contrast to traditional approaches where all fermion degrees of freedom in the entire thermodynamic volume are summed over in order to solve the sign problem. In some four-fermion models, the degrees of freedom inside a fermion bag often involve only a small fraction of the total number of degrees of freedom in the entire thermodynamic volume and can be summed up quickly yielding a positive weight. This makes the fermion bag approach very powerful. It is an extension of the meron cluster idea proposed some time ago \cite{Chandrasekharan:1999cm}.

In order to illustrate the idea consider models formulated with $n$ tastes of massless staggered fermions. These models contain $2n$ Grassmann variables per site which will be denoted as $\psi_i(x)$ and $\psib_i(x)$ where $i=1,2,..n$ represent taste indices and $x$ denotes the space time lattice point. Let $D$ be the $V \times V$ free staggered Dirac matrix whose matrix elements are denoted as $D_{xy}$. The properties of $D$ are such that any $k$-point correlation function of chiral condensates
\begin{equation}
C_i(x_1,...,x_k) = \int [d\psib d\psi] \exp\Big(\sum_{x,y} \ \psib_i(x) \ D_{xy}\ \psi_i(y) \Big) \psib_{i}(x_1)\psi_{i}(x_1)\ ...\ \psib_{i}(x_k)\psi_{i}(x_k)
\end{equation}
involving the taste $i$ is always positive. Using Wick contractions it is easy to prove that
\begin{equation}
C_i(x_1,..,x_k) = \mathrm{Det}(D)\ \ \mathrm{Det}(G[\{x\}])
\label{smallu}
\end{equation}
where $G[\{x\}]$ is the $k \times k$ matrix of propagators between the $k$ sites $x_i,i=1,..,k$ whose matrix elements are $G_{x_i,x_j} = D^{-1}_{x_i,x_j}$. It is also possible to argue that \cite{Chandrasekharan:2009wc},
\begin{equation}
C_i(x_1,..,x_k) = \mathrm{Det}(W[\{x\}])
\label{largeu}
\end{equation}
where the matrix $W[\{x\}]$ is the same as the matrix $D$ except that that the sites $\{x\} \equiv \{x_i,i=1,2,..k\}$ are dropped from the matrix. Thus, $W$ is a $(V-k) \times (V-k)$ matrix. The identity 
\begin{equation}
\mathrm{Det}(D)\ \ \mathrm{Det}(G[\{x\}]) \ =\ \mathrm{Det}(W[\{x\}])
\label{duality}
\end{equation}
is the basis behind the weak coupling-strong coupling duality we mentioned in the previous section.

Now consider a generic four fermion lattice field theory action involving $n$ massless staggered fermions given by
\begin{equation}
S = - \sum_{x,y,i } \ \psib_i(x) \ D_{xy} \psi_i (y) - \sum_{\langle xy \rangle} \sum_{i,j} \ U_{i,j,\langle xy \rangle} 
\psib_i(x)\psi_i(x)  \psib_j(y)\psi_j(y)  
\end{equation}
where $\langle xy\rangle$ refers to some well defined set of neighboring lattice sites. Further we will assume that all the couplings $U_{i,j,\langle xy\rangle }$ are non-negative real constants. The partition function of the model 
\begin{equation}
Z = \int \ [d\psib d\psi] \  \mathrm{e}^{-S}
\end{equation}
can be expanded in powers of the couplings $U_{i,j,\langle xy\rangle}$ and each term in the expansion consists of a product of correlation functions of the type $C_i(x_1,...,x_k)$ introduced above. Mathematically the expansion looks like
\begin{equation}
Z = \sum_{[\{x\}]} \{[U]\}^p \ \prod_{i=1}^n C_i(x_1,...,x_{k_i})
\end{equation}
where $\{[U]\}^p$ refers to some generic power of the coupling and $k_i,i=1,2,..n$ refers to the number interaction vertices for each taste. Note that on a finite lattice the expansion is a very high order polynomial and so no convergence issues arise.

 An intuitive physical picture emerges from the above expansion of the partition function. Since the $k_i$ interaction sites contain both $\psi_i$ and $\psib_i$, fermions are already paired on these sites and do not cause sign problems\footnote{Any remaining sign problem would also be present in a bosonic field theory.}. On the other hand, unpaired fermions that hop freely on the remaining sites can indeed cause sign problems. The free sites are collectively referred to as a fermion bag. The summation of all fermion world lines inside the bag should be a determinant of a $(V-k_i) \times (V-k_i)$ matrix. Indeed $C_i(x_1,x_2,..,x_k) = \mathrm{Det}(W[\{x\}])$. 
The determinant can be evaluated easily if $(V-k_i)$ is small, which naturally occurs at strong couplings. Hence we refer to these bags as strong coupling fermion bags. It was shown in \cite{Chandrasekharan:2009wc} that at strong couplings a fermion bag splits into many small disconnected pieces making things even simpler. The left figure of Fig.~\ref{fig:bag} gives an illustration of the disconnected pieces of a fermion bag at strong couplings. The solid bonds represent interaction sites while the sites in the shaded regions are free sites and form the fermion bag. The arrows inside the shaded regions is an illustration of free fermion world lines inside the bag.

Clearly at weak couplings the number of free sites grows enormously. Thus, the definition of a fermion bag as a set of free sites, which was natural at strong couplings loses its charm at weak couplings. However, thanks to a duality we can construct the fermion bag differently. At weak couplings we can view the interactions as the unpaired fermionic degrees of freedom that cause fluctuations over the paired fermionic free vacuum as they hop from one interaction site to another. The effort to compute these fermionic fluctuations should only grow as the determinant of a $k_i \times k_i$ matrix. The duality relation of Eq.(\ref{duality}) shows that $C_i(x_1,x_2,..,x_k) = \mathrm{Det}(W[\{x\}]) = \mathrm{Det}(D)\ \ \mathrm{Det}(G[\{x\}])$ where the fluctuation matrix $G[\{x\}]$ is indeed a $k_i \times k_i$ matrix. The right figure of Fig.\ref{fig:bag} gives an illustration of a weak coupling fermion bag. The solid bonds again represent interaction sites which form the fermion bag. Fermions hop from one interaction site to another through the free fermion propagator. One set of hopping is shown by arrows in the figure.

We must acknowledge that the weak coupling fermion bag idea is exactly equivalent to the idea of summing over all Feynman diagrams and was introduced earlier in the framework of diagrammatic determinantal Monte Carlo methods. For a review please see Ref.~\cite{RevModPhys.83.349}. Indeed the sum over all Wick contractions at the given order in perturbation theory is simply $\mbox{Det}(G[\{x\}])$. On the other hand, we think the fermion bag approach as a more natural interpretation at least in the context of lattice field theories since it uncovers the powerful concept of duality discussed above and shows that the approach is efficient both at weak and strong couplings.

The fermion bag approach is rather general and is equally applicable to non-relativistic fermions and models with Wilson fermions. However, sometimes it is necessary to introduce unconventional interactions like eight fermion interactions. In some models, the weight of a fermion bag is no longer a determinant but involves new mathematical structures like fermionants \cite{Chandrasekharan:2011an}. These are very similar to determinants, and may be exponentially difficult to compute \cite{MertensS2011}. For these models, the fermion bag approach is not practical. 

\section{Massless Thirring Model}

As a first application of the fermion bag approach we have studied the three dimensional massless lattice Thirring model. It is a lattice field theory containing massless fermions and an interesting quantum critical point. Variants of this model have been used recently to describe the critical points associated with Graphene~\cite{PhysRevB.84.075123}. The model is constructed with two Grassmann valued fields $\psi(x)$ and $\psib(x)$ on each site $x$ of a cubic lattice. The lattice action is given by
\begin{equation}
S = -\sum_{x,y} \ \psib(x) \ D_{xy}\ \psi(y) - U \sum_{\langle x y\rangle}  \ 
\psib(x) \psi(x) \ \psib(y)\psi(y).
\label{eq:model}
\end{equation}
Here $\langle xy\rangle$ refers to the set of nearest neighbor sites across a bond. We use anti-periodic boundary conditions in all the three directions. The four fermion coupling $U$ generates the current-current coupling of the Thirring model in the continuum. The lattice model is invariant under a $U_f(1)\times U_\chi(1)$ symmetry, where $U_f(1)$ is the fermion number symmetry and $U_\chi(1)$ is the chiral symmetry. When the coupling $U$ is small the model contains $N_f=2$ flavors of massless four-component Dirac fermions at long distances due to fermion doubling. At large $U$ the chiral symmetry breaks spontaneously and generates a single massless Goldstone boson while the fermions become massive. There is a quantum critical point $U_c$ which separates the phase with massless fermions from the phase with massless bosons. 

The model has been studied earlier using mean field techniques~\cite{PhysRevLett.59.14}, and conventional Monte Carlo methods~\cite{DelDebbio:1995zc,Debbio:1997dv,DelDebbio:1997hd,Barbour:1998yc}. In particular properties of the quantum critical point have been computed. However, all previous work was done in the presence of a fermion mass and on lattice sizes which are not very big compared to the correlation lengths introduced due to the presence of a fermion mass. Thus, the analysis may contain uncontrolled systematic errors. A fresh Monte Carlo method which works directly in the massless limit should be useful. This is what we accomplish here using the fermion bag approach. 

The fermion bag approach for the model was first developed in~\cite{Chandrasekharan:2009wc}. However, in the previous study only strong coupling bags were used since the concept of duality was not appreciated. Here we repeat the study with weak coupling bags and are able to push the study to larger lattice sizes. In the fermion bag approach the partition function of the model in Eq.(\ref{eq:model}) can be rewritten as
\begin{equation}
Z = \sum_{[n]} U^{N_b} \mbox{Det}(W[n])
\end{equation}
where $[n]$ refers to the configuration of interaction bonds $n_{\langle xy\rangle} = 0,1$ and $N_b$ refers to the total number of bonds. For a given configuration $[n]$, the free sites form the strong coupling fermion bag (see the left figure in Fig.~\ref{fig:bag}). However this representation of the partition function is not useful when $N_b$ is a small fraction of the volume. For example, at the quantum critical point $N_b$ is about an eighth of the lattice volume. On the other hand, one can use the duality relation
\begin{equation}
\mbox{Det}(W[n]) = \mbox{Det}(D)[\mbox{Det}^2(G[n])]
\end{equation}
to simplify the computation when $N_b$ is small. Here $D$ is the free staggered fermion matrix on the whole lattice and $G[n]$ is the $N_b \times N_b$ free fermion propagator matrix between even lattice sites to the odd lattice sites of the bonds. This representation is equivalent to a redefinition of the fermion bag as the set of interaction sites (see the right figure of Fig.~\ref{fig:bag}). 

\section{Observables}

In order to uncover the properties of the quantum critical point we have measured four observables:
\begin{enumerate}
\item The simplest is the average number of bonds $\langle N_b \rangle$. Fluctuations in this observable allow us to monitor equilibration and autocorrelation times easily.
\item Since we work with exactly massless fermions the chiral condensate will always be zero. However, the chiral condensate susceptibility is nonzero. It is defined as
\beq
\chi = \frac{1}{2 L^3}\sum_{x,y}\langle\psib_x\psi_x\psib_y\psi_y\rangle.
\eeq
We expect $\chi$ to scale as $L^{2-\eta}$ at the critical point. A related quantity is the bosonic two-point function:
\beq
C_B(|x-y|)=\langle \psib_x\psi_x\psib_y\psi_y\rangle
\eeq
We will define the ratio $C_B(L/2-1)/C_B(1)$ as a useful way to track autocorrelations.
\item Another useful observable that measures the onset of chiral symmetry breaking is the chiral winding susceptibility $\langle q^2_\chi \rangle$. If we define the conserved chiral charge passing through the surface $S$ perpendicular to the direction $\alpha$ as
\beq
(q_\chi)_\alpha=\sum_{x \in S} \ \varepsilon_x\ \eta_{x,\alpha}\ (D^{-1})_{x,x+\alpha}\ +\ \sum_{x\in S}\  2\varepsilon_x,
\eeq
where  $\varepsilon_x = (−1)^{x_1+x_2+x_3}$ is the parity of the site x, then $\langle q^2_\chi \rangle=\langle \frac{1}{3}\sum_\alpha(q^2_\chi)_\alpha \rangle$, is expected to be independent of $L$ at the critical point. This allows us to determine the critical point accurately.
\item  Since one of the novel features of the current quantum critical point is the presence of massless fermions at the critical point, the fermion two-point correlation function is expected to show non-trivial scaling. Hence we define
\beq
C_F(d)=\frac{1}{3}\sum_{\alpha = 1}^3 \langle\bar{\psi}_x\psi_{x+d\hat{\alpha}}\rangle
\eeq
where $x$ belongs to a site with $\varepsilon(x) = 1$ and $\hat{\alpha}$ is a unit vector along each of the three directions. We compute the ratio $R_f=C_F(L/2-1)/C_F(1)$ which is expected to scale as $L^{-(2+\eta_\psi)}$ at the critical point.
\end{enumerate}

\begin{figure}
\begin{center}
\includegraphics[width=3.0in]{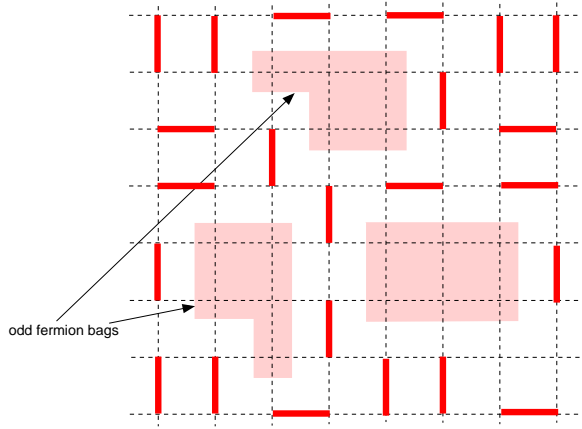}
\caption{An illustration of a zero weight configuration due to the presence of a fermion bag with an odd number of free sites. The staggered fermion Dirac operator defined on an odd number of lattice sites has an exact zero mode}\label{fig:zero}
\end{center}
\end{figure}

\section{The Algorithm}

We now discuss the details of the Monte Carlo algorithm we have used to solve the model. An important technical problem is that some observables like $\chi$ get contributions from configurations that do not contribute to the partition function. For example, the configuration  illustrated in Fig.~\ref{fig:zero} contributes to $\chi$, but has zero weight in the partition function. This is because, in the massless limit any strong coupling fermion bag with an odd number of free sites has zero weight. However, introducing a $\psib(x)\psi(x)$ source in each of the two odd bags makes the number of free sites even and thereby changing the configuration weight to be nonzero. Hence instead of computing $\chi$ (which can be done in principle \cite{Chandrasekharan:2009wc}) we redefine our partition function to be 
\beq
Z_2 = \sum_{x,y} \int \ d\psib d\psi \ (\psib_x\psi_x \ \psib_y\psi_y) \ \mathrm{e}^{-S}.
\eeq
which always contains two sources. We then redefine $\chi \equiv 1/f_2$ where $f_2$ is the fraction of the configurations which contain sources one lattice unit apart. The redefined $\chi$ has all the same finite size scaling properties as the original definition of $\chi$ close to the quantum critical point. Below we will describe the Monte Carlo algorithm for simulating the partition function $Z_2$. The algorithm for simulating the real partition function will be the subset of updates in which the sources can be ignored.

The algorithm consists of  four updates: (1) Bond creation/destruction, (2) Bond translation, (3) Source translation, (4) Worm update. Each update begins with a configuration $[n]$ where $2d$ bond variables $n_{x,\alpha}=0,1$ at each site $x$ are completely defined. Note that $\alpha=\pm 1,\pm 2,\cdots\pm d$ denote the directions (the negative signs indicate negative directions). If a bond exists between $x$ to $x+\hat{\alpha}$ then $n_{x,\alpha}=1$.  In our notation $n_{x,\alpha} \equiv n_{x+\hat{\alpha},-\alpha}$.  The configuration $[n]$ also defines source variables $m_x = 0,1$ such that $m_x = 1$ on only two sites in the entire lattice. On these sites $n_{x,\alpha} = 0$. The parity of the site $x$ is given by $\varepsilon_x = (−1)^{x_1+x_2+\cdots+x_d}$.

\subsection{Bond Creation/Destruction}

Bond creation and destruction is accomplished through a simple Metropolis algorithm. The exact update is as  follows:
\begin{enumerate}
\item With probability $1/2$ we propose to destroy a bond. If $N_b=0$ the update stops. Otherwise we pick one of bonds randomly with probability $1/N_b$. We destroy that bond with probability
\beq
P (N_b\rightarrow N_b-1)=\frac{N_b}{N_f U}\frac{W^\prime}{W}
\eeq
where $W=\mbox{Det}^2\Big\lbrace G([n])\Big\rbrace$ and $W^\prime = \mbox{Det}^2\Big\lbrace G([n^\prime])\Big\rbrace$ where $[n^\prime]$ is the new configuration obtained by destroying the bond. $N_f$ denotes the number of the locations where we would have been allowed to create a bond starting from the configuration $[n^\prime]$.

\item With probability $1/2$ we decide to create a bond. Let $N_f$ to be the number of locations where we could create a bond. If $N_f=0$ the update stops. Otherwise we pick one of $N_f$ locations with probability $1/N_f$. We then create the bond at that location with probability
\beq
P(N_b \rightarrow N_b+1)=\frac{N_f U}{N_b}\frac{W^\prime}{W}
\eeq
where $W=\mbox{Det}^2\Big\lbrace G([n])\Big\rbrace$ and $W^\prime = \mbox{Det}^2\Big\lbrace G([n^\prime])\Big\rbrace$  where $[n^\prime]$ is the new configuration obtained by creating the new bond.
\end{enumerate}
If we begin with a configuration with a small number of bonds, since $N_f \gg N_b$ the acceptance probability $P(N_b \rightarrow N_b+1)$ is close to one and  the algorithm creates bonds efficiently.  Once $N_f U$ is of the order of $N_b$, the system begins to thermalize and the average number bonds $\langle N_b\rangle$ fluctuates only a little. 

\subsection{Bond Translation}

In order to reduce autocorrelation times it is useful to have an update which can move bonds from one location to another. This update is again a Metropolis update and is meaningful only when $N_b \neq 0$. We pick an existing bond at random, destroy it and create a bond at another allowed location picked at random with probability
\beq
P=\frac{W^\prime[n^\prime]}{W[n]}
\eeq
where $W^\prime[n^\prime]$ and $W[n]$ are the weights of the new configuration and old configurations.

\subsection{Source Translation}
We need to update the locations of the two sources in the configurations that contribute to $Z_2$.  We again accomplish this using a Metropolis algorithm. We pick one of the sources at random, destroy it and create it at an allowed site with the same parity as the site where the source was destroyed, with probability
\beq
P=\frac{W^\prime[n^\prime]}{W[n]}
\eeq
where $W^\prime[n^\prime]$ and $W[n]$ are the weights of the new and old configurations.

\subsection{Worm Updates}

Configurations with two sources come in two varieties. One set of configurations contain both sources in a single strong coupling fermion bag, while the two sources appear in two different strong coupling fermion bags in the other set. The above three updates cannot change between these two set of configurations easily and this can lead to a long autocorrelation time. The reason is that it is impossible to remove the source from a bag and move it to another bag in one step. Only a series of correlated moves can accomplish this task. The problem is most severe in the broken phase and perhaps close to the quantum critical point. Fortunately, worm updates of the type discussed in Ref.~\cite{Adams:2003cc} completely eliminate this problem. Further, they are extremely fast since they do not require the computation of any determinants. We perform two types of worm updates : 
\begin{enumerate}
\item {\bf Bond Update}
\begin{enumerate}
\item We  pick a site $x$ at random and define it as the first site. 
\item If $n_{x,\alpha} = 0$, the update ends. Otherwise, we break the bond by creating a source at $x$ and $x+\hat{\alpha}$ (i.e., set $n_{x,\alpha}=0$, $m_x = 1$ and  $m_{x+\hat{\alpha}}$) and move to the site $y = x+\hat{\alpha}$. 
\item We then pick a direction $\mu$ such that $x' = y+\hat{\mu}$ contains a bond ($n_{x',\alpha'} = 1$) or $x'$ is the first site ($x'=x$). We pick this direction at random from the available choices.
\item If $n_{x',\hat{\alpha'}} = 1$, then we break that bond and  create a bond that connects the site $y$ and $x$ and move the source from $y$ to $x'+\hat{\alpha'}$ (i.e., we set $n_{y,\mu} = 1$, $m_y = 0$ and $m_{x'+\hat{\alpha'}}=1$). We then redefine $y=x'+\hat{\alpha'}$  and go back to the previous step. 
\item If $x'=x$, then we remove the two sources at $y$ and $x$ and create a bond connecting the two sites (i.e. set $m_y=0$, $m_x=0$ and $n_{y,\mu} = 1$). Then the update ends. Note that the update begins and ends at the same site and hence is a loop update
\end{enumerate}
\item {\bf Source Update}
\begin{enumerate}
\item We pick one of the two sources at site $x$.
\item We look for all directions $\mu$ that contains a bond. If no such site exists the update ends. Otherwise 
a direction is picked at random from the available choices. 
\item If the site $x+\hat{\mu}$ contains a bond in the $\alpha$ direction we break this bond and create a new bond that connects $x$ and $x+\hat{\mu}$ and move the source at site $x$ to the site $x+\hat{\mu}+ \hat{\alpha}$ (i.e., set $m_x = 0$, $m_{x+\hat{\mu}+\hat{\alpha}} = 1$ and $n_{x,\mu} = 1$). The update then ends. 
\end{enumerate}
\end{enumerate}
When the source update is repeated many times it can move the source from one bag to another bag. The source update is very cheap and can be repeated thousands of times within a second.

\subsection{Determinant Computation}

Except for the worm updates, all other updates require the computation of the determinant of $G[n]$ which is an $N_b \times N_b$ matrix where $N_b$ is the number of bonds. A change in $[n]$ by a single bond or source location typically changes one row and one column of this matrix. The most time consuming part of the algorithm is calculating the determinants before and after the change. Changes in $[n]$ can lead to singular matrices quite often. Further, sub-matrices of $G[n]$ can also be singular. There are ways to compute the determinant of a matrix quickly if a single row and column are changed \cite{PhysRevB.72.035122}. However, these tricks involve inverses and must be thought through carefully due to possible singularities.

In our work we use LU decomposition with complete pivoting to compute the determinant. However, this approach requires $O(N_b^3)$ computations. Further, a single sweep requires roughly $N_b$ bond updates. Hence, naively the computations necessary to accomplish a single sweep will scale like $O(N_b^4)$. However, as we explain below, we can speed up the update by a factor of $N_b$, thus reducing the number of computations for a single sweep to $O(N_b^3)$.

The basic idea is that if one decides apriori that a fixed set of $l$ rows and $l$ columns will be updated in the matrix $G[n]$, then while performing the LU decomposition we can organize our calculations so that we can reuse a large part of the computation. For example, consider the LU decomposition of the block matrix $G[n]$
\beq
G[n]=\left(
\begin{array}{cc}
A&B\\
C&D
\end{array}
\right)
\label{eq:block_matrix}
\eeq
where $A$, $B$, $C$ and $D$ are respectively $N_b-l \times N_b - l$, $N_b-l\times l$, $l \times N_b-l$ and $l \times l$ matrices. If the matrix $A$ is not singular, we can compute the determinant of $G[n]$ as
\beq
\mbox{Det} G[n] = \mbox{Det}A \ \cdot\ \mbox{Det} (D - C A^{-1}B)
\label{eq:block_det}
\eeq
Since $A$ is fixed during the update, the determinant of $G[n]$ due to varying the elements in $B, C, D$ requires only $O(l N_b^2)$ computations. If $A$ is singular, then while computing the $LU$ decomposition of $A$ we can isolate the zero mode sector and merge it with the part that is being updated. Since the zero mode sector of $A$ is always tiny the above approach works well. 

In order to implement the above idea, we divide the full volume into $N_{\rm blocks}$ such that every bond belongs to a unique block. Each block contains roughly $l$ bonds. We then perform the three time consuming updates on the bonds associated to a randomly chosen block $j$. We choose a basis such that the matrix $G[n]$ can be written as a block matrix as shown in Eq.~(\ref{eq:block_matrix}).  In particular the matrix $A$ contains propagators only between sites of bonds that do not belong to the block $j$, while the matrix $D$ contains only propagators between sites of the bonds in $j$. The matrices $B$ and $C$ contain propagators between the block $j$ and outside. We then compute the LU decomposition with full pivoting on the matrix $A$. This allows us to isolate any singular part of the matrix $A$ if it exists and merge it with the matrices $B,C$ and $D$.  Since the singular part also does not change, it can be easily taken into account if necessary. The updates described above can be adapted to each block by redefining $N_b$ and $N_f$ as numbers obtained within the block $j$. By choosing $N_{block}\sim O(\sqrt{N_b})$, the number of computations for one sweep scales like $O(N_b^3)$. We believe our method is similar to the ``fast-updates'' algorithm of the usual determinantal algorithm.

\begin{figure}[h]
\begin{center}
\includegraphics[width=3.5in]{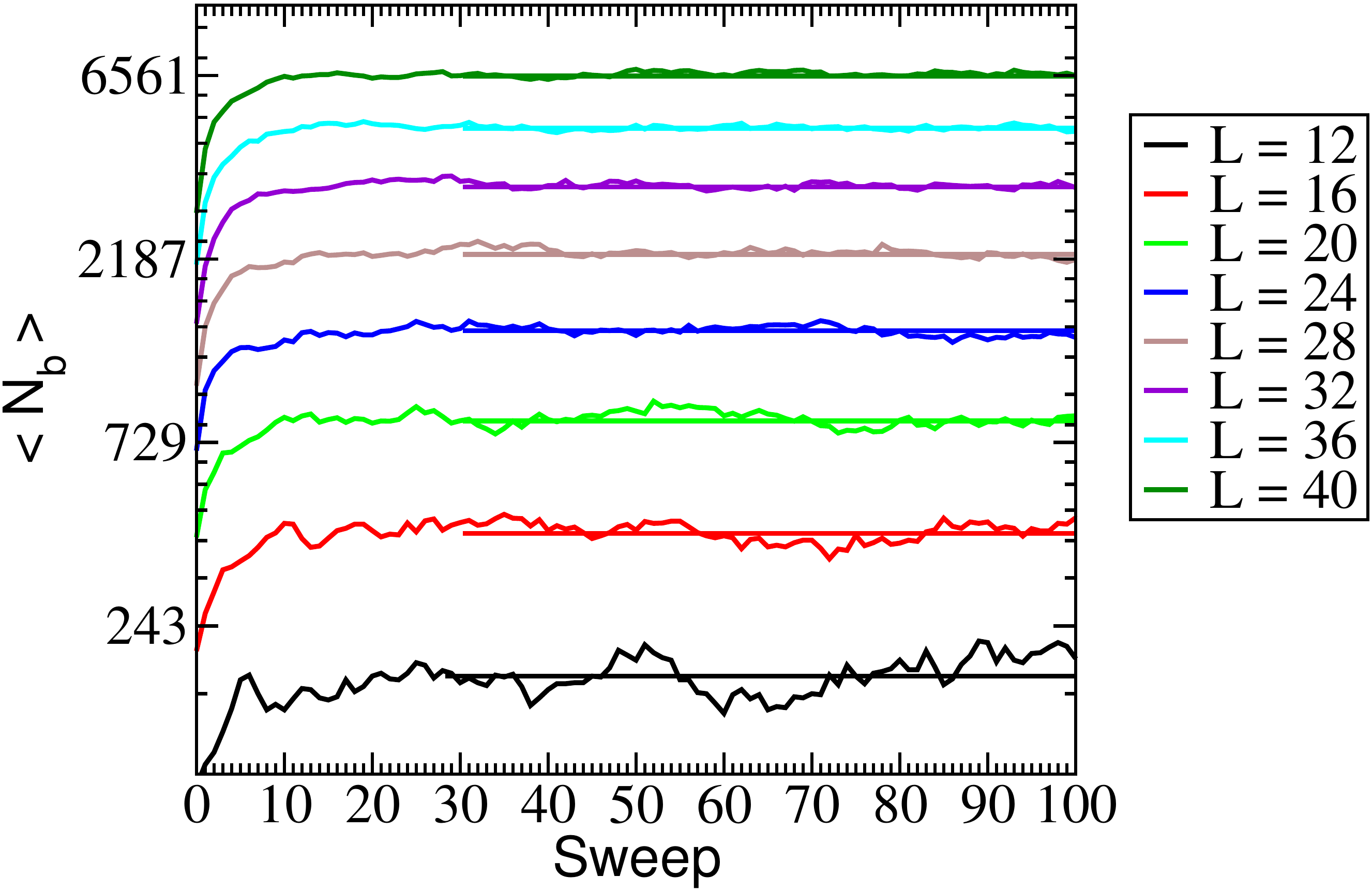}
\caption{Thermalization time for $N_b$ in terms of sweep using Log plot}
\end{center}
\label{fig:therm}
\end{figure}

\begin{figure}[h]
\begin{center}
\includegraphics[width=2.9in]{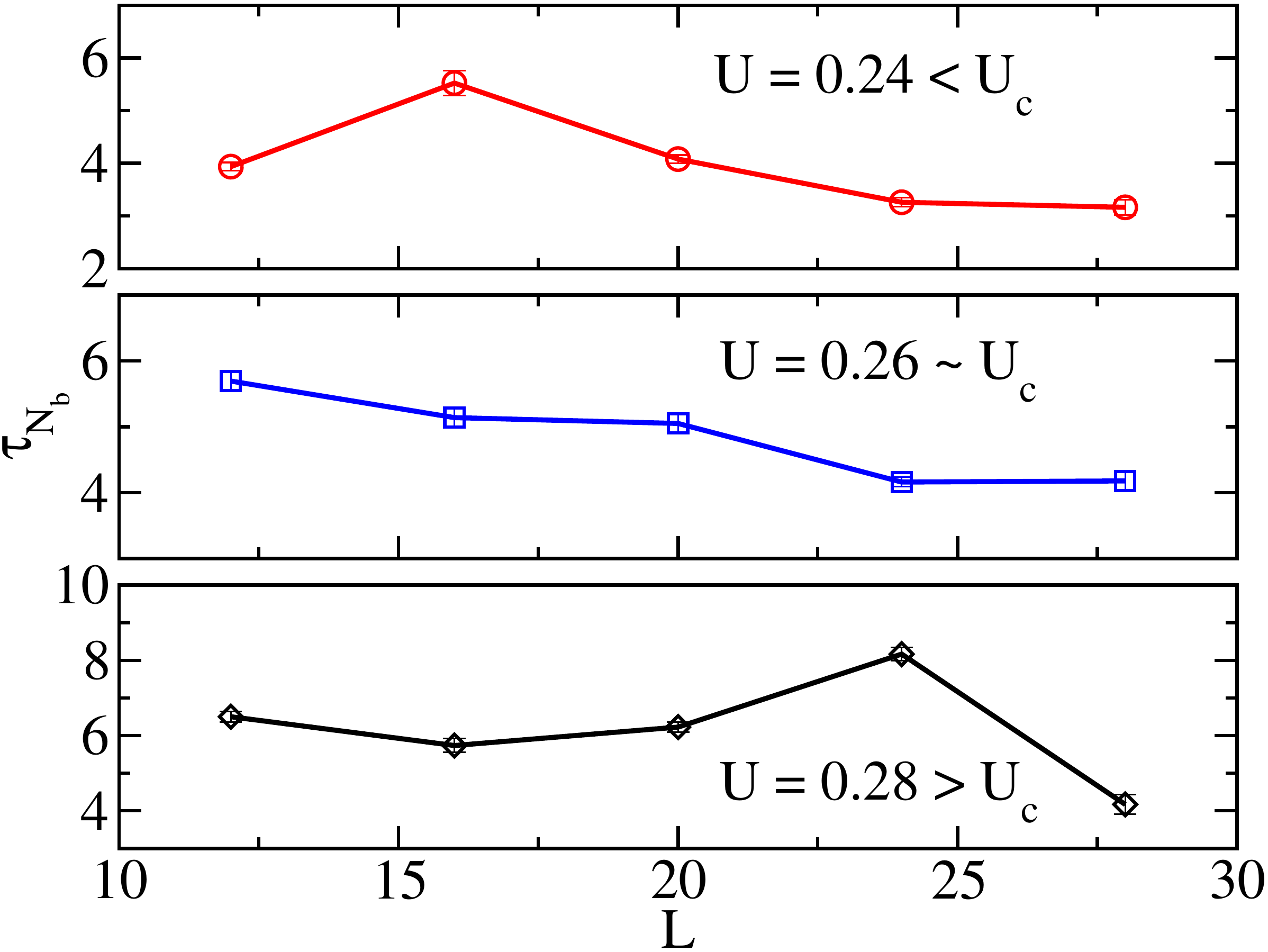}
\includegraphics[width=3in]{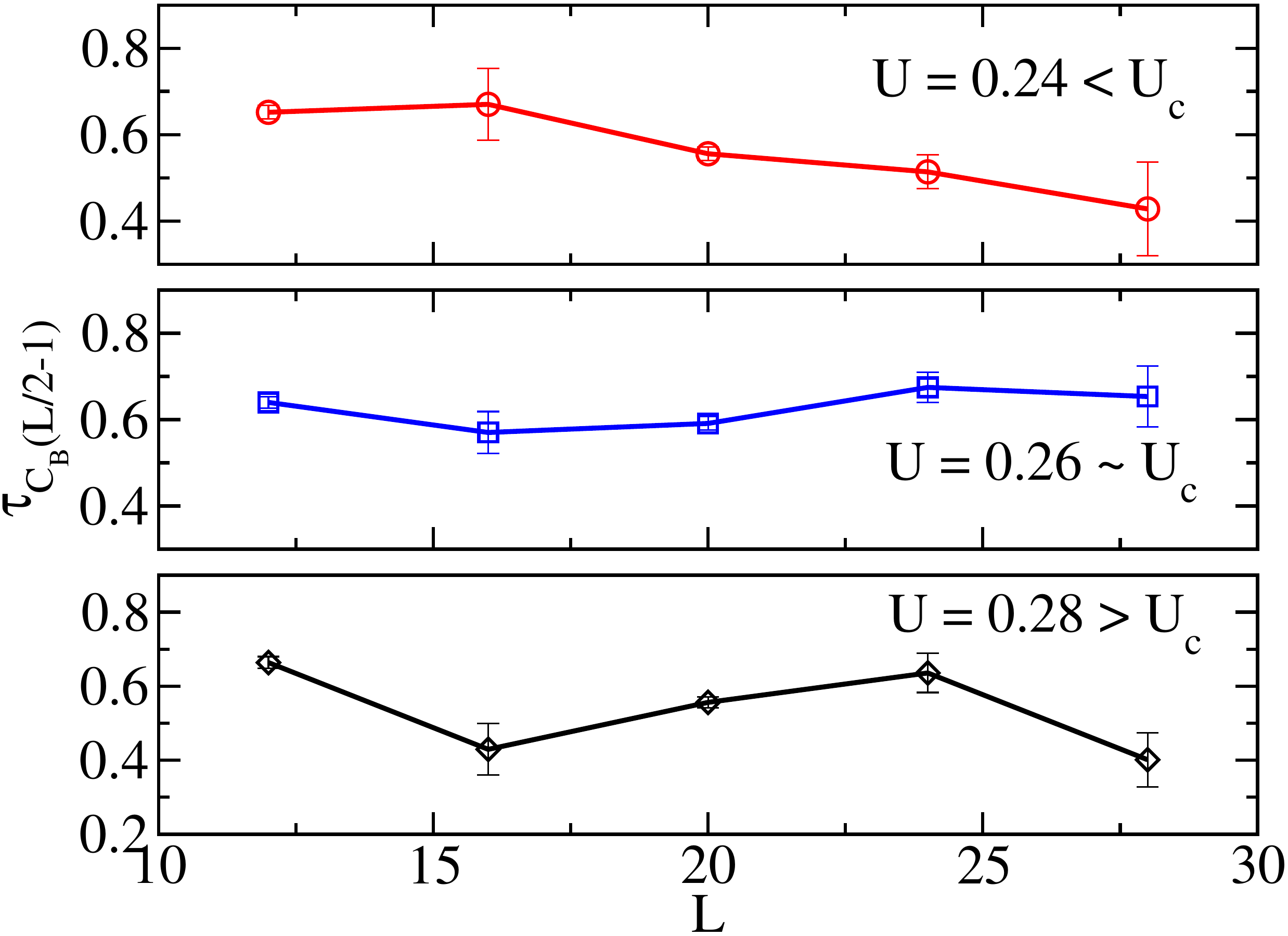}
\caption{Autocorrelation time for $N_b$ (Left) and $C_B$ (Right) at $|x-y|=L/2-1$ in terms of sweep. It shows the autocorrelation time does not depend on $L$ even at $U_c$}
\end{center}
\label{fig:auto}
\end{figure}

\subsection{Performance of the Algorithm}

We have studied thermalization times and autocorrelation times for two observables, $\langle N_b\rangle$ and $C_B(L/2-1)$, in units of a sweep. In our work, a sweep is defined as $V/8$ updates of all types discussed above \footnote{Since worm updates are cheap we perform several thousand worm updates per sweep.}. In Fig~\ref{fig:therm}, we plot the thermalization time in units of a sweep, for $\langle N_b\rangle$ for different volumes where we initialize configurations with $V/24$ randomly chosen bonds. The figure shows the system at different volumes can be thermalized within roughly the same number of sweeps. In Fig~\ref{fig:auto}, we plot the autocorrelation time $\tau$ in units of a sweep. One typically expects $\tau \propto L^z$, where $z$ is the dynamical critical exponent of the algorithm. For most local algorithms $1\leqslant z \leqslant 2$. Many efficient cluster algorithms on the other hand are known to have $0 < z \leqslant 1$. Surprisingly, our data shows that $z \approx 0$ even at $U = U_c$. However, we must emphasize that the time spent for each sweep grows as $(fV)^3$ where $f < 1$ is some fixed fraction at a given value of $U$. At the critical point $f \sim 1/8$. Despite this bad scaling with the volume, since $f$ is small we have been able to obtain results on lattices as large as $40^3$ with modest computing resources.

\begin{figure}[t!]
\begin{center}
\vbox{
\hbox{
\includegraphics[width=2.8in]{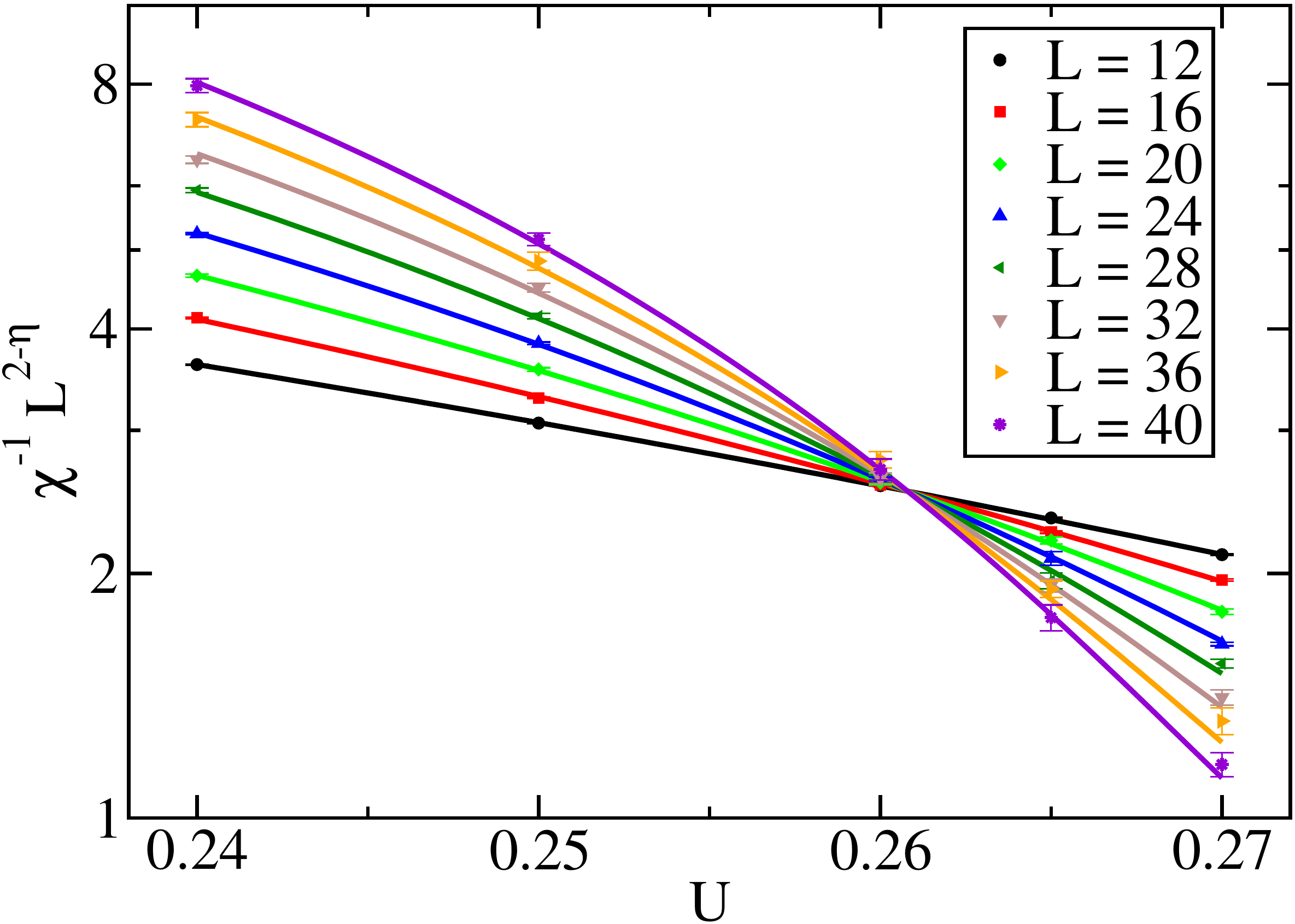}
\hskip0.1in
\includegraphics[width=2.8in]{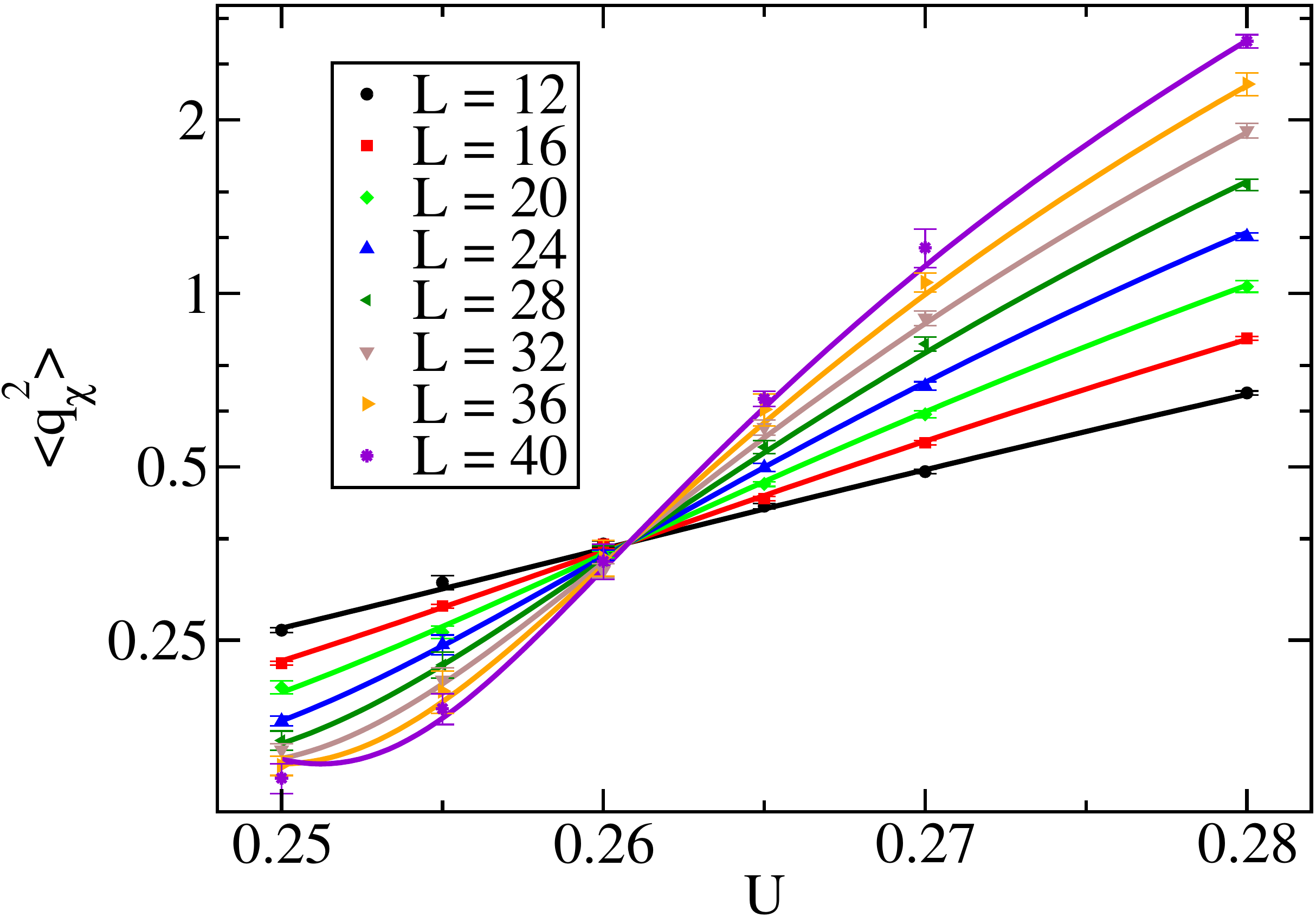}
}
\vskip0.1in
\hbox{
\hskip1.3in
\includegraphics[width=2.9in]{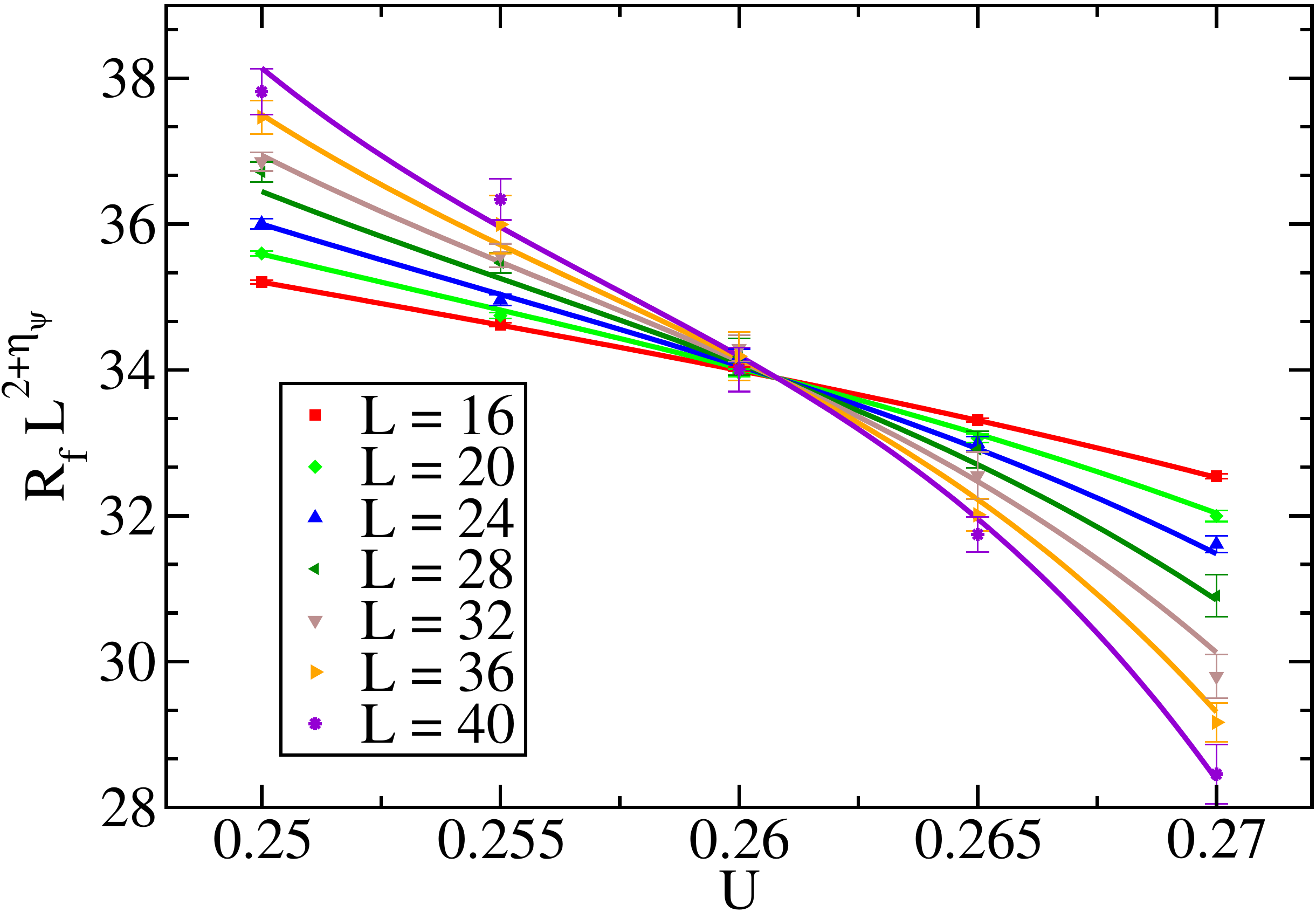}
}
}
\caption{Plots of $\chi^{-1} L^{2-\eta}$, $\langle q_\chi^2\rangle$ and $R_f L^{2+\eta_\psi}$ as a function of $U$ for $L = 12, 16, 20, 24, 28, 32, 36$ and $40$. The solid lines show the combined fit. Based on the fits we find the critical point to be $U = 0.2608(2)$, and three critical exponents are $\nu = 0.85(1), \eta=0.65(1)$ and $\eta_\psi=0.37(1)$.}
\label{fig:scaling}
\end{center}
\end{figure}

\section{Results}

One of the main goals of our work is to compute the critical exponents at the quantum critical point. These have been calculated earlier in Refs.~\cite{DelDebbio:1995zc,Debbio:1997dv,DelDebbio:1997hd, Barbour:1998yc} using the HMC method with the traditional approach in which the four-fermion interaction is converted to a fermion bilinear with the help of an auxiliary field. However, these calculations were performed in the presence of a quark masses and on lattice volumes that were not very big. The presence of two infrared scales, namely the fermion mass and the length of the box can be difficult to take into account in the finite size scaling relations which can lead to large systematic errors. In our work, since the fermions are exactly massless, the analysis is much simpler and cleaner. We focus on $\chi$, $\langle q^2_\chi \rangle$ and $R_f$ in the vicinity of $U_c$ where we expect the following finite size scaling relations to hold:
\begin{subequations}
\label{eq:scaling}
\beqa
\chi^{-1} L^{2-\eta} = \sum_{k=0}^3 f_k \left[(U-U_c) L^{\frac{1}{\nu}}\right]^k \\
\langle q_\chi^2 \rangle = \sum_{k=0}^3 \kappa_k \left[(U-U_c) L^{\frac{1}{\nu}}\right]^k \\
R_f L^{2+\eta_\psi} = \sum_{k=0}^3 p_k \left[(U-U_c) L^{\frac{1}{\nu}}\right]^k 
\eeqa
\end{subequations}
For each observable, when the left hand side is plotted as a function of $U$, curves for different values of $L$ must cross at $U=U_c$. A combined fit to all the Eqs.~(\ref{eq:scaling}) gives the critical exponents $\nu = 0.85(1), \eta=0.65(1)$ and $\eta_\psi=0.37(1)$ and $U = 0.2608(2)$ with a  $\chi^2/d.o.f.=1.3$. The complete list of fit parameters are listed in Table~\ref{table:fit}. The results from the combined fit are plotted in Fig.~\ref{fig:scaling}. For comparison, one of the earlier work finds $\nu = 0.71(4)$ and $\nu=0.60(2)$ \cite{DelDebbio:1997hd}.

\begin{table*}[h]
\caption{Results for the fit parameters from the combined fit of the data to Eqs.~(6.1).}
\begin{center}
\begin{tabular}{c c c c c c c c}
\hline\hline
$\eta$ & $\eta_\psi$ & $\nu$ & $U_c$ &$\kappa_0$ & $\kappa_1$& $\kappa_2$& $\kappa_3$\\
\hline
0.65(1) & 0.37(1) & 0.85(1) & 0.2608(2) & 0.369(3) & 0.63(1) & 0.52(2) & 0.09(1) \\
\hline\hline
$f_0$ & $f_1$& $f_2$& $f_3$ & $p_0$ & $p_1$& $p_2$& $p_3$\\
\hline
2.52(3) & -2.53(5) & 0.71(3) & 0.10(1) & 33.9(2) & -5.0(1) & -2.0(2) & -2.5(5)\\
\hline\hline
\label{table:fit}
\end{tabular} 
\end{center}
\end{table*}

\section{Conclusions}

In this work we have shown that the recently proposed fermion bag approach is a powerful technique for solving some four-fermion lattice field theories. Due to an interesting duality, the approach is efficient both at weak and strong couplings and continues to perform well at intermediate couplings. As a first application of the method we studied the critical behavior in the massless Thirring model and found an algorithm that practically eliminates critical slowing down when times are measured in units of sweep. The time to perform a sweep scales as $O(N_b^3)$ where $N_b$ is the size of a fermion bag which is usually a small fraction of the volume. In the massless thirring model we found that $N_b \sim V/8$ at the quantum critical point. The Hybrid Monte Carlo method has never been successfully applied to massless fermions on large volumes.  We have accomplished this using the fermion bag approach on lattices as large as $40^3$ with moderate computing resources. We were able to compute the critical exponents at the quantum critical point.

\section*{Acknowledgments}

We thank Philippe deForcrand, Simon Hands, David Kaplan, Keh-Fei Liu, Costos Strouthos and Uwe-Jens Wiese for discussions. This work was supported in part by the Department of Energy grants DE-FG02-05ER41368 and  DE-FG02-00ER41132.

\end{document}